\documentclass[preprint,showpacs,preprintnumbers,amsmath,amssymb,aps,pre]{revtex4-1}
\usepackage{graphicx}
\usepackage{dcolumn}
\usepackage{bm}
\usepackage[shortlabels]{enumitem}
\begin{document}
.
\preprint{LA-UR-16-26145}

\title{V-T theory for the Self-Intermediate Scattering Function in a Monatomic Liquid}
\author{Duane C. Wallace}
\affiliation{Theoretical Division, Los Alamos National Laboratory, 
Los Alamos, New Mexico 87545}\author{Eric D. Chisolm}
\affiliation{Theoretical Division, Los Alamos National Laboratory, 
Los Alamos, New Mexico 87545}
\author{Giulia De Lorenzi-Venneri}
\affiliation{Theoretical Division, Los Alamos National Laboratory, 
Los Alamos, New Mexico 87545}

\date{\today}
\begin{abstract}
In V-T theory the atomic motion is harmonic vibrations in a liquid-specific potential energy valley, plus transits, which move the system rapidly among the multitude of such valleys. In its first application to the self intermediate scattering function  (SISF), V-T theory produced an accurate account of molecular dynamics (MD) data at all wave numbers $q$ and time $t$. Recently, analysis of the mean square displacement (MSD) resolved a crossover behavior that was not observed in the SISF study. Our purpose here is to apply the more accurate MSD calibration to the SISF, and assess the results. We derive and discuss the theoretical equations for vibrational and transit contributions to the SISF. The time evolution is divided into three successive intervals: the vibrational interval when the vibrational contribution alone accurately accounts for the MD data; the crossover when the vibrational contribution saturates and the transit contribution becomes resolved; and the diffusive interval when the transit contribution alone accurately accounts for the MD data. The resulting theoretical error is extremely small at all $q$ and $t$. Comparison of V-T and mode-coupling theories for the MSD and SISF reveals that, while their formulations differ substantially, their underlying atomic motions are in logical correspondence.

\end{abstract}

\maketitle

\section{Introduction}
V-T theory of liquid dynamics is being developed within the framework of many-body theory. The liquid atomic motion has two contributions, many-body vibrational motion described by a first-principles liquid-specific Hamiltonian, and transit motion described by a parameterized model. Vibrations express the dominant part of the liquid thermal energy, while transits provide the liquid diffusion. It was recently shown that a proper adjustment of the transit parameters yields an extremely accurate account of molecular dynamics (MD) data for the mean square displacement time-correlation function (MSD) for liquid Na~\cite{MSD}. In the present study we employ precisely the same atomic motion to calculate the self-intermediate scattering function (SISF) of the same liquid Na system. While the MSD constitutes a single scalar measure of the motional decorrelation process, the SISF measures the complete Fourier transform of that process. This extra complication of the SISF, plus the use of an independent calibration of the motion, poses a stringent test of the V-T theoretical formulation. Our aim here is to carry out this test by comparing theory with MD, and assess the results. 

Unless otherwise stated, we consider monatomic systems. By hypothesis, the potential surface underlying the liquid atomic motion is overwhelmingly dominated  by the random class of $3N$-dimensional potential energy valleys~\cite{DCW1997}. These valleys are macroscopically uniform, so that a single such valley is sufficient for statistical mechanical calculations~\cite{DCW1997}. This hypothesis has been verified extensively for the present Na system~\cite{WC1999, CW1999, Holm2009, Holm2010}, and has been verified by density functional theory (DFT) calculations for Na and Cu~\cite{Nick2010}, for Ga~\cite{Sven2014}, and for a five component metallic glass~\cite{Holm2010}. The hypothesis is verified in each instance where a single-valley vibrational contribution rationalizes MD data, as in the MSD graphs~\cite{MSD}, and in the SISF graphs of the present study. A more detailed discussion of the liquid potential energy surface, and of the technique for calculating the vibrational Hamiltonian, is provided in~\cite{MSD}.

Ninety years ago,  Frenkel argued that a liquid atom oscillates for a time about one equilibrium position, then jumps to a new one and oscillates there~\cite{F1926, F1946}. In developing V-T theory, we have added much crucial information to this description. First, the vibrational motion is now the fully correlated $3N$-dimensional normal-mode motion. Second, transits replace jumps. Individual transits were observed in low-temperature equilibrium MD trajectories, in metastable liquid states of Ar and Na~\cite{WCC2001}. Each observed transit is the highly correlated motion of a small local group of atoms that carries the system across the boundary between two liquid potential energy valleys. These transits are mechanical motion, not comparable to the much larger statistical-mechanical cooperatively rearranging regions of Adam and Gibbs~\cite{AG1965}. The transit-induced motion of the atomic equilibrium positions, abbreviated transit motion, is continuous and has physically meaningful time dependence (section~III; see also~\cite{MSD}).Transit theory is still under development, and the present work is a step in that direction.

In section~II, vibrational theory for the SISF is outlined, and the complete system correlation components and decorrelation processes are mathematically defined. Transit theory for the SISF is outlined in section~III, and the transit damping coefficients are derived. Equations for the theoretical SISF are derived in section~IV, and their accuracy is verified for a representative $q$ value. In section~V, the complete theory is applied to a set of $q$ spanning all physical behaviors, from hydrodynamic at small $q$ to free particle at large $q$.  The deviation of the theory from MD data is tabulated and discussed. Mode coupling theory (MCT) and V-T theory are compared in terms of their respective descriptions of the underlying atomic motion in section~VI. Major conclusions are outlined in section~VII.

\section{Vibrational theory}

We study a system representing liquid Na at $395$K, based on a well-tested interatomic potential from pseudopotential perturbation theory (see figure 1 in~\cite{WC1999}). The system has $N$ atoms in a cubical box with periodic boundary conditions, with $N=500$ here. The atomic positions are $\bm{r}_{K}(t)$ at time $t, K=1,\dots,N$, and $\{\bm{q}\}$ is the set of wavevectors commensurate with the box periodicity. The SISF is~\cite{HMcD_3rded,BZ1994}
\begin{equation} \label{eq1}
F^{s}(q,t) = \frac {1}{N}   \left < \sum_{K} e^{-i \bm{q}\cdot (\bm{r}_{K}(t)-\bm{r}_{K}(0))}                                        
            \right >.
\end{equation}
The brackets indicate the average over the atomic motion in an equilibrium state, plus the average over all $\bm{q}$ for each $q$ magnitude. The $\bm{q}$ average is the last to be done, and we shall omit its explicit notation. We work in classical statistical mechanics, which allows comparison with MD data and which is accurate for most elemental liquids, including Na. Equation~(\ref{eq1}) is used for direct calculation of the MD function 
$F_{MD}^{s}(q,t)$.

Vibrational theory for the SISF is derived and discussed in~\cite{G1}; a brief summary is helpful here. 
The vibrational contribution to the liquid atomic motion is harmonic vibrations in a single random valley. Each atom moves with displacement $ \bm{u}_{K}(t)$ away from the fixed equilibrium position $\bm{R}_{K}$, so that
\begin{equation} \label{eq2}
\bm{r}_{K}(t)=\bm{R}_{K} + \bm{u}_{K}(t). 
\end{equation}
For this motion, equation~(\ref{eq1}) is
\begin{equation} \label{eq3}
F_{vib}^{s}(q,t)=\frac {1}{N}   \left < \sum_{K} e^{-i \bm{q}\cdot (\bm{u}_{K}(t)-\bm{u}_{K}(0))}                                       
            \right >_{vib},
\end{equation}
where $<\dots>_{vib}$ indicates the average over the vibrational motion. With Bloch's identity (equation N15 of \cite{AM1976}), equation~(\ref{eq3}) becomes
\begin{equation} \label{eq4}
F_{vib}^{s}(q,t)=\frac {1}{N}  \sum_{K}  e^{-2W_{K}(\bm{q})}\;
     e^{\left < \bm{q}\cdot\bm{u}_{K}(t) \;\;\bm{q}\cdot\bm{u}_{K}(0)\right >_{vib}},
\end{equation}
where
\begin{equation} \label{eq5}
2 W_{K}(\bm{q})=\left < (\bm{q}\cdot\bm{u}_{K}(0))^{2}\right > _{vib}.
\end{equation}

To evaluate the average in equation~(\ref{eq4}), the vibrational Hamiltonian is diagonalized in terms of the vibrational normal modes, labeled $\lambda=1,\dots,3N-3$, where the three zero frequency modes are omitted from all calculations. Following the algebra in section 5.1 of \cite{CW2001}, we find
\begin{equation} \label{eq6}
\left < \bm{q}\cdot\bm{u}_{K}(t)\;\; \bm{q}\cdot\bm{u}_{K}(0)\right >_{vib} =
\frac{kT}{M} \sum_{\lambda} (\bm{q}\cdot \bm{w}_{K\lambda})^{2}
\; \frac{\cos \omega_{\lambda}t}{\omega_{\lambda}^{2}},
\end{equation}
where $T$ is temperature, $M$ is the atomic mass, $\bm{w}_{K\lambda}$ is the Cartesian vector of the $K$ component of eigenvector $\lambda$, and  $\omega_{\lambda}$ is the corresponding frequency. The functions $e^{-2W_{K}(\bm{q})}$ in equation~(\ref{eq4}) are Debye-Waller factors, and the left side of equation~(\ref{eq6}) is the vibrational time-correlation function.

\begin{figure} [h]
\includegraphics [width=0.70\textwidth]{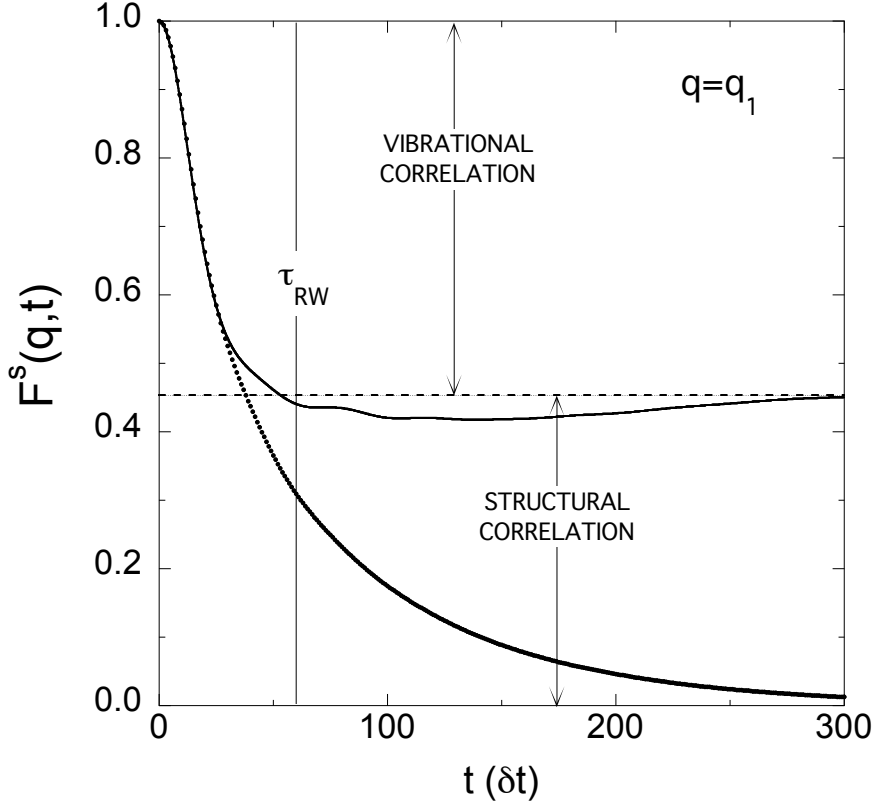}
\caption{Dots show $F_{MD}^{s}(q,t)$ at MD time steps, and line is $F_{vib}^{s}(q,t)$. Dashed line is $F_{vib}^{s}(q,\infty)$, the $t\rightarrow \infty$ limit of $F_{vib}^{s}(q,t)$. The system correlation per atom at $t=0$ has the marked vibrational and structural components. $\tau_{RW}$ is explained in the text.}
\label{fig1}
\end{figure}

In this section and the next two, we shall illustrate the theory for a single representative $q$, namely $q_{1}$ at the first peak of the structure factor $S(q)$. Figure~\ref{fig1} shows $F_{MD}^{s}(q,t)$ and $F_{vib}^{s}(q,t)$, together with the ultimate constant value $F_{vib}^{s}(q,\infty)$ of $F_{vib}^{s}(q,t)$. 
At $t=0$, the contribution to equation~(\ref{eq1}) is $1$ for every atom, and this is maintained in equations~(\ref{eq4}) and (\ref{eq5}). As $t$ increases from zero, the theoretical motion is free particle (ballistic), for which $\bm{r}_{K}(t)-\bm{r}_{K}(0)=\dot{\bm{r}}_{K}(0) t$. With this motion the leading $t$-dependence of equation~(\ref{eq1}) is
\begin{equation} \label{eq7}
F_{free}^{s}(q,t) = e^{-a(q)t^{2}},
\end{equation}
where $a(q)=kTq^{2}/2M$. Precisely this leading $t$-dependence, to exponential order, is contained in equations~(\ref{eq4}) and (\ref{eq6}). It is shown below that the vibrational motion is pure ballistic for only a very short time.

Following the ballistic regime, the $t$-dependence of $F_{vib}^{s}(q,t)$ is controlled by vibrational dephasing, which is the interference of the $\cos \omega_{\lambda}t$ terms in equation~(\ref{eq6}). This process continues until the vibrational time correlation functions are zero. Then from equation~(\ref{eq4}), the ultimate value of $F_{vib}^{s}(q,t)$ is
\begin{equation} \label{eq8}
F_{vib}^{s}(q,\infty) = \frac{1}{N} \sum_{K}e^{-2W_{K}(\bm{q})}, \;\; t\geq t_{\infty}.
\end{equation}
From extension of figure~\ref{fig1} to longer times we find $t_{\infty}\approx 300 \delta{t}$. $t_\infty$ is not a calibration time in the present study.

We can define the correlations and the decorrelation processes involved in the SISF at all $q$ and $t$. The correlations are between $\bm{r}_{K}(t)$ and $\bm{r}_{K}(0)$, averaged over the motion, as expressed in equation~(\ref{eq1}). In V-T theory, these correlations have two independent components, vibrational and structural. At $t=0$, the vibrational correlation  is $1-F_{vib}^{s}(q,\infty)$ per atom, and the structural correlation is $F_{vib}^{s}(q,\infty)$. This decomposition is marked in figure~\ref{fig1}.

\begin{figure} [h]
\includegraphics [width=0.70\textwidth]{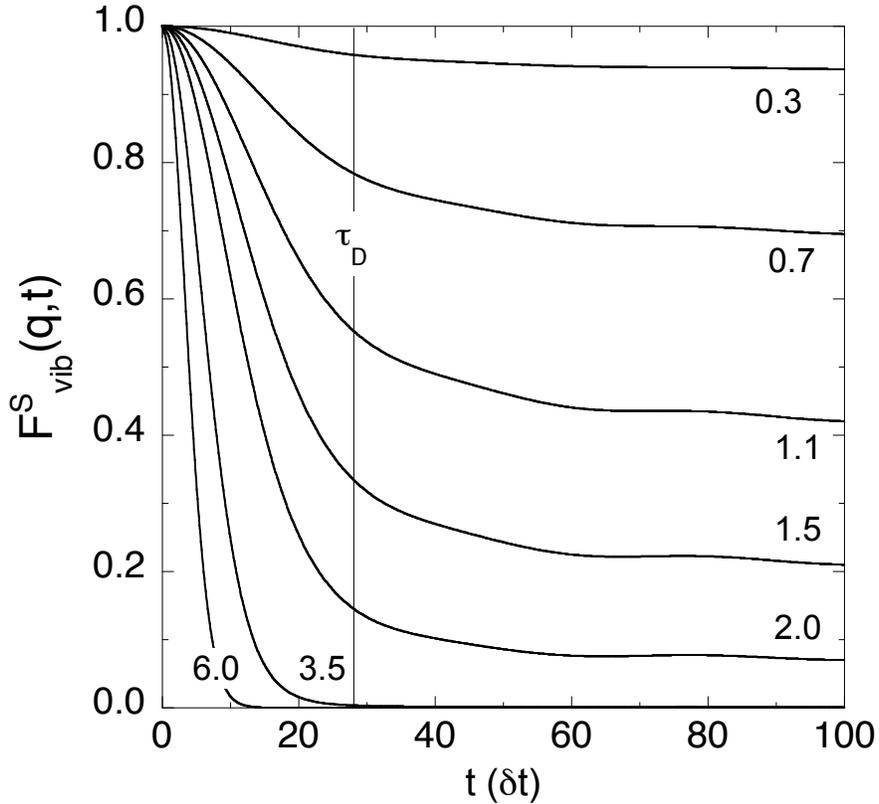}
\caption{For $q$ from $0.3$ to $6.0$  a$_{0}^{-1}$: $F_{vib}^{s}(q,\infty)$ is near $1$ at $q=0.3$a$_{0}^{-1}$ and goes to zero around $\tau_{D}$ at $q=3.5$.  As $q$ increases from $3.5$, $F_{vib}^{s}(q,t)$ goes to zero in an ever shorter time.}
\label{fig2}
\end{figure}

The Debye-Waller factor measures the mean structural correlation for a system of atoms vibrating about fixed equilibrium positions. For such a system, e.g. an amorphous solid, the structural correlation is  not damped. Vibrational dephasing damps the vibrational correlation to zero while the structural correlation remains. This is shown by $F_{vib}^{s}(q,t)$ in figure~\ref{fig1}. The $q$-dependence of this process is controlled by the $q^2$ factor in equation~(\ref{eq5}), which appears in the exponent of equation~(\ref{eq8}). This makes $F_{vib}^{s}(q,\infty)$ approach 1 as $q\rightarrow 0$, and carries $F_{vib}^{s}(q,\infty)$ to zero as $q$ increases. The progression is shown in figure~\ref{fig2}. The representative $q$ chosen here is $q_{1}=1.1$, which divides the initial correlation nearly evenly between vibrational and structural components (see figure~\ref{fig1}). The primary function of transits in the present theory is to damp the structural correlation to zero.

\section{Transit Theory}

\begin{figure} [h]
\includegraphics [width=0.70\textwidth]{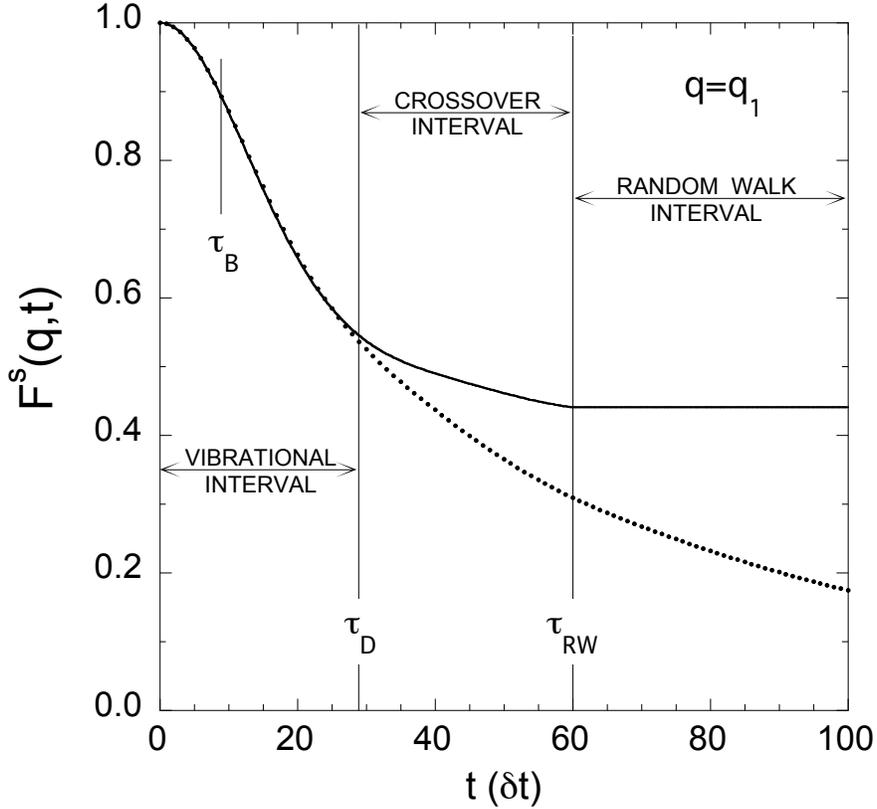}
\caption{Dots are $F_{MD}^{s}(q,t)$, line is $F_{vib}^{s}(q,t)$, and their  difference is due to transit damping. The slope discontinuity in $F_{vib}^{s}(q,t)$ at $\tau_{RW}$ is due to a small transit effect. $\tau_{B}$ is explained in the text.}
\label{fig3}
\end{figure}

Figure~\ref{fig3} compares  $F_{MD}^{s}(q,t)$  and $F_{vib}^{s}(q,t)$ on three time intervals. The curves in figure~\ref{fig3} approximate an inversion of the corresponding MSD curves~\cite{MSD}. Our MSD study provides the following brief summary of the atomic motion underlying figure~\ref{fig3}. On the first interval, until the delay time $\tau_{D}$, the MD system measures only pure vibrational motion. On the crossover interval, the vibrational contribution saturates to its ultimate constant value while the MD curve continues to decay toward zero. The difference between the MD and  vibrational curves is transit damping, which is first resolved at $\tau_{D}$ and reaches its full steady state at $\tau_{RW}$. Besides damping the structural correlation, an additional small transit effect is to damp away the final vibrational correlation, and make  $F_{vib}^{s}(q,t)$ constant at $t \geq \tau_{RW}$. The approximation for this effect introduces the small slope discontinuity in $F_{vib}^{s}(q,t)$ at $\tau_{RW}$ (for details see~\cite{MSD}). 

The commonly studied time correlation functions, including the ones treated here, measure the trajectory of one atom at a time, as in equation~(\ref{eq1}). We therefore need only the separate transit motion of a single atom, without explicit accounting of its motional correlation with other atoms. The single atom behavior is encoded in the V-T decomposition of the liquid motion:
\begin{equation} \label{eq9}
\bm{r}_{K}(t)=[\bm{R}_{K}(t)-\bm{R}_{K}(0)] + [\bm{u}_{K}(t)+\bm{R}_{K}(0)]. 
\end{equation}
The second bracket is vibrational motion about the permanent equilibrium position $\bm{R}_{K}(0)$, as in equation~(\ref{eq2}). The first bracket expresses the transit induced motion of the atomic equilibrium positions. The distance moved by the equilibrium position of one atom in one transit, averaged over transits, is denoted $\delta{R}$.  The motion $\delta{R(t)}$ starts from zero, and the liquid system cannot sense the transit motion until it has reached a sufficient magnitude. The corresponding time required for the MD system in figure~\ref{fig3} to resolve a transit, and begin measuring it, is identified as $\tau_{D}$. This explains why the MD system measures only vibrational motion for $t$ up to $\tau_{D}$.  

Transits are proceeding uniformly throughout the liquid at a high rate. The one-atom transit rate is $\nu$, and we make the following uniform transit timing approximation. Every atom transits once in the transit period $\nu^{-1}$, the transits are uniformly distributed over time, and the same sequence of one-atom transits occurs in every succeeding transit period. In an increment of time, a time correlation function averages this motion over a great many transits.

By some time after $t=0$, every transit motion is resolved from its beginning and measured to its completion. Then the net effect of the transit motion is to move the equilibrium position of every atom a randomly directed distance $\delta{R}$ in a time $\nu^{-1}$. This motion constitutes a steady-state random walk of each atomic equilibrium position. It is no longer necessary to account for the $t$-dependence of the transit motion; we can simply record each transit as it is completed.

The damping factor for the steady-state transit random walk is derived in~\cite{G1}. Now we need damping factors for two successive time intervals. To set this up, we consider $\bm{R}_{K}(t)-\bm{R}_{K}(0)$, from equation~(\ref{eq9}), to be a random walk of step rate equal to the transit rate $\nu$, but of arbitrary step distance $S$ and start time $\tau$. Equation~(\ref{eq1}) then takes the form
\begin{equation} \label{eq10}
F_{VT}^{s}(q,t) = \frac {1}{N}   \sum_{K}  \left < \left <e^{-i \bm{q}\cdot (\bm{R}_{K}(t)-\bm{R}_{K}(0))}\;\;e^{-i \bm{q}\cdot (\bm{u}_{K}(t)-\bm{u}_{K}(0))}                                        
            \right >_{trans} \right>_{vib},
\end{equation}
where $<\dots>_{trans}$ is the average over transits. We shall neglect correlations between the two motions, and average the first exponential uniformly over all system transits. This average removes the $K$ dependence of the first factor and also removes its dependence on the direction of $\bm{q}$. The required average can therefore be written
\begin{equation} \label{eq11}
A(q,t)=\left <e^{-i \bm{q}\cdot (\bm{R}(t)-\bm{R}(0))}   \right >_{trans}.
\end{equation}
Equation~(\ref{eq10}) becomes
\begin{equation} \label{eq12}
F_{VT}^{s}(q,t)=A(q,t)\; F_{vib}^{s}(q,t)
\end{equation}
where $F_{vib}^{s}(q,t)$ is given by equations~(\ref{eq3}) and (\ref{eq4}).

In a time increment $\delta{t}$, we have
\begin{equation} \label{eq13}
\delta A(t )= \left< \left[e^{-i \bm{q}\cdot \bm{R}(t+\delta t)}-e^{-i\bm{q}\cdot \bm{R}(t)}\right]
e^{i \bm{q}\cdot\bm{R}(0)}  \right>_{trans}.
\end{equation}
We shall record each transit step in the $\delta{t}$ in which it is completed. The probability a given atom completes a step in $\delta{t}$ is $\nu \delta{t}$. If the atom completes a step in $\delta{t}$,  $\bm{R}(t+\delta t) = \bm{R}(t) + \bm{S}$; otherwise  $\bm{R}(t+\delta t) = \bm{R}(t)$. Therefore
\begin{equation} \label{eq14}
\delta A(t)=\left< [ e^{-i \bm{q}\cdot \bm{S}}-1]
\;\;e^{-i\bm{q}\cdot (\bm{R}(t)-\bm{R}(0))} \right>_{trans} \nu \delta t.
\end{equation}
The average over transits is now an average over the uniformly distributed directions of $\bm{S}$. Equation~(\ref{eq14}) becomes
\begin{equation} \label{eq15}
\frac{\delta A(q,t)}{\delta t}= - \gamma(qS)\;A(q,t),
\end{equation}
where
\begin{equation} \label{eq16}
\gamma(x) = \nu \left[1-\frac{\sin x}{x}\right].
\end{equation}
Since the random walk begins at $\tau$, the solution of equation~(\ref{eq15}) is
\begin{equation} \label{eq17}
 A(q,t) = \begin{cases}
  1                   &\text{for $t\leq \tau$,} \\
 e^{-\gamma(qS)(t-\tau)} &\text{for $t\geq \tau$.} 
\end{cases}
\end{equation}
The above equations will be assembled as needed to produce transit damping in the crossover and random walk intervals.

\section{Calibration and verification of theory}

We begin with the early calibration of two transit parameters, $\nu$ and $\delta{R}$. In our study of the MD trajectories of transiting atoms~\cite{WCC2001}, we recorded the mean  $\delta{R}$ as around $0.40 R_{1}$ for Ar and $0.25 R_{1}$ for Na, where $R_{1}$ is the nearest neighbor distance. We have since used this  $\delta{R}$ for liquid Na near the melting temperature $T_{m}$, since  $\delta{R}$ is expected to have little $T$-dependence. The study was not exhaustive, and  $\delta{R}$ can be in error by $25\%$ or so.

In order to calibrate the theoretical $F_{VT}^{s}(q,t)$, $\nu$ and $\delta{R}$ are needed separately in $\gamma(q\delta{R})$ (see equation~(\ref{eq16}). From an earlier calculation of the self-diffusion coefficient $D$ for liquid Na, we used the Einstein relation $D=\frac{1}{6} \nu(\delta{R})^2$, along with the above value of  $\delta{R}$, to find $\nu$~\cite{G1}. The parameter combination $\nu(\delta{R})^2$  is extremely accurate, since the evaluation of $D$ is accurate to $1\%$ for our liquid Na system. These parameter calibrations, along with the others described below, are reported in~\cite{MSD} and are listed in Table I. We shall now discuss the remaining calibrations and the verification of theory for each interval in turn.

\begin{table}[ht] 
\caption{\label{table1}Values for the parameters as defined in the text. The MD time step is $\delta t = 7.00288$~fs.}
\begin{ruledtabular}
\begin{tabular}{ccccc}
$\nu$&$\delta R$&$S$&$\tau_{D}$&$\tau_{RW}$\\
\hline
3.9~ps$^{-1}$&1.75~a$_{0}$&1.46~a$_{0}$&$28~\delta t$&$60~\delta t$\\
\end {tabular}
\end{ruledtabular}
\label{Table1}
\end{table}

\subsection{Vibrational interval:}

\begin{figure} [h]
\includegraphics [width=0.70\textwidth]{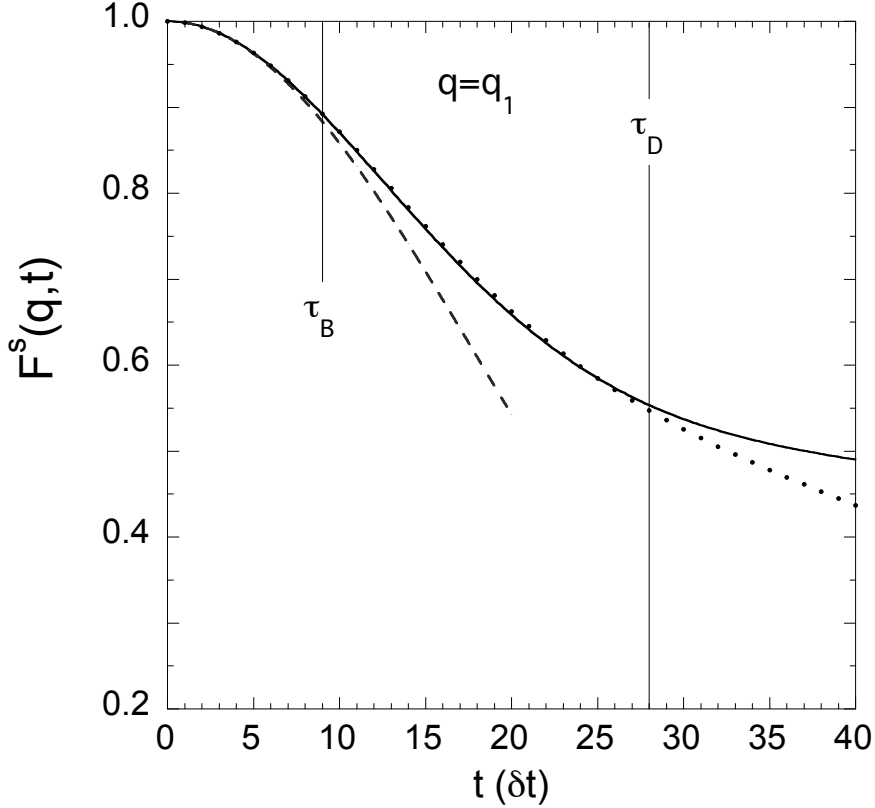}
\caption{Dots are $F_{MD}^{s}(q,t)$, line is $F_{vib}^{s}(q,t)$, and dashed line is the ballistic contribution, which is accurate only to $\tau_{B}=9\delta{t}$.}
\label{fig4}
\end{figure}
In our MSD study, $\tau_{D}$ is defined operationally as the time when $X_{MD}(t)$ begins to move away from  $X_{vib}(t)$, and $\tau_{D}$ is calibrated from the comparison graph of the two functions. To apply the MSD calibration to  $F_{VT}^{s}(q,t)$, we write
\begin{equation} \label{eq18}
F_{VT}^{s}(q,t)=F_{vib}^{s}(q,t), \;\;  0\leq t\leq \tau_{D}.
\end{equation}
The comparison with $F_{MD}^{s}(q,t)$ in figure~\ref{fig4} shows excellent agreement for the function and the time interval. Other details identical between MSD and SISF are: a) The ballistic contribution, shown in figure~\ref{fig4}, is accurate only to $\tau_{B}=9\delta{t}$, and accounts for only a very small part of the vibrational decorrelation; b) In figure~\ref{fig4}, $F_{VT}^{s}(q,t)$ shows a tiny positive departure from $F_{MD}^{s}(q,t)$, through the center of the interval, similar to the negative MSD departure in figure 2 of~\cite{MSD}, and negligible in both cases.

\subsection{Crossover interval:}

\begin{figure} [h]
\includegraphics [width=0.70\textwidth]{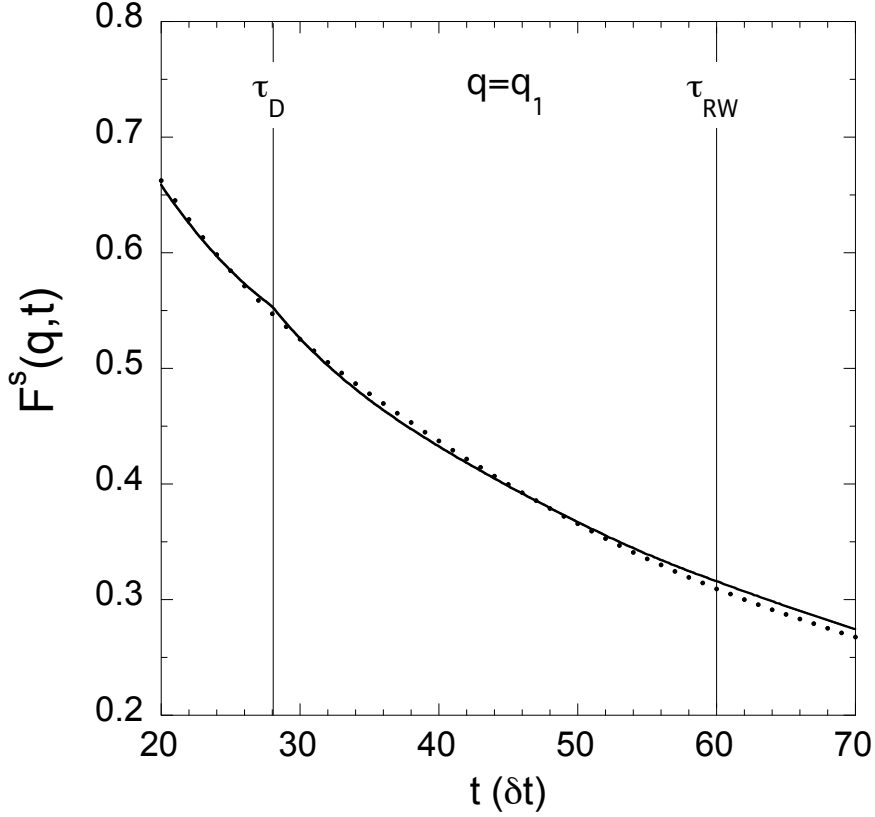}
\caption{Dots are $F_{MD}^{s}(q,t)$ and line is $F_{VT}^{s}(q,t)$ up to $\tau_{RW}$. The line beyond $\tau_{RW}$ is a continuation of crossover theory.}
\label{fig5}
\end{figure}
On the crossover interval, the MSD transit contribution is very well fitted by a straight-line segment from $\tau_{D}$ to $\tau_{RW}$ (figure (4) of~\cite{MSD}). Because it is linear in $t$, the straight line represents a random walk. However, this random walk is incomplete because its step distance $S$ is less than the complete $\delta{R}$ (see Table I). To calibrate 
$F_{VT}^{s}(q,t)$ we must employ the same ``crossover walk" expressed in equation (10) of ~\cite{MSD}. For a random walk of step rate $\nu$, step distance $S$, and starting time $\tau$, $F_{VT}^{s}(q,t)$ is given by equations~(\ref{eq12}), (\ref{eq16}) and (\ref{eq17}). The present result for the  crossover walk is therefore
\begin{equation} \label{eq19}
F_{VT}^{s}(q,t)=F_{vib}^{s}(q,t) \;\;e^{-\gamma(qS)(t-\tau_{D})}       , \;\;  \tau_{D}\leq t\leq \tau_{RW}.
\end{equation}
Comparison with $F_{MD}^{s}(q,t)$ in figure~\ref{fig5} shows excellent agreement.

Going back to the MSD, it is possible to achieve better agreement between theory and MD on the crossover. However, we chose to fit the MSD crossover with a simple random walk, because only that motion through  equations~(\ref{eq10})-(\ref{eq17}) leads to the physically simple and tractable damping solution on the right side of  equation~(\ref{eq19}).

\subsection{Random walk interval:}

\begin{figure} [h]
\includegraphics [width=0.70\textwidth]{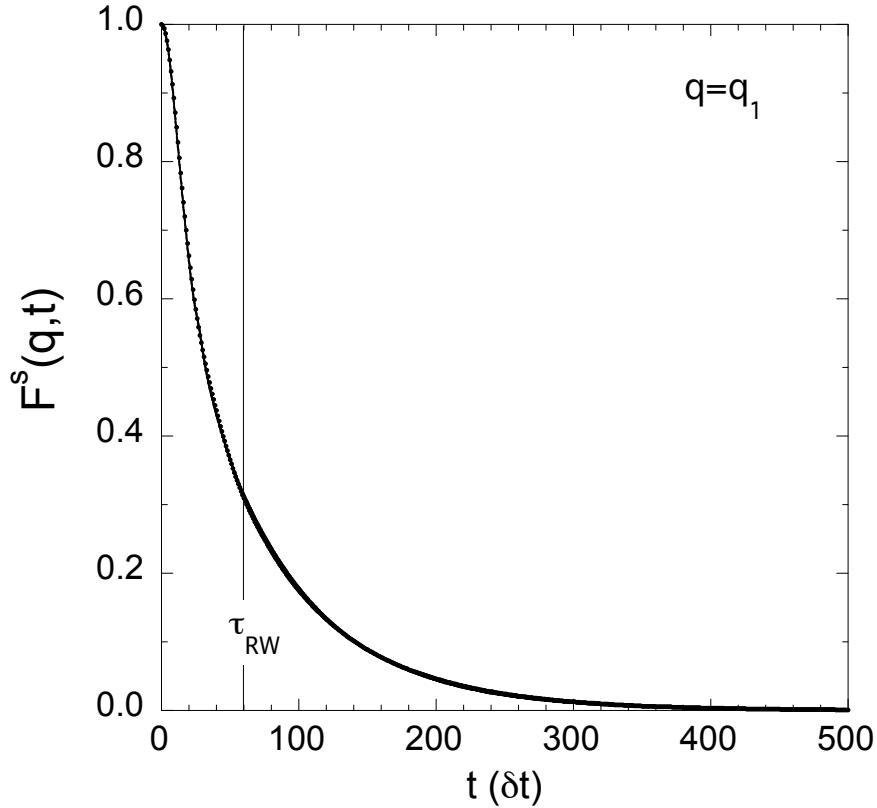}
\caption{Dots are $F_{MD}^{s}(q,t)$ and  line is $F_{VT}^{s}(q,t)$. The deviation is insignificant at all $t$.}
\label{fig6}
\end{figure}

In our MSD study, $\tau_{RW}$ is defined operationally as when the mean square displacement, $X_{MD}(t)$, begins its ultimate linear-in-$t$ straight line, whose slope is $6D=\nu(\delta{R})^{2}$. The calibration of $\tau_{RW}$ is possible because $X_{MD}(t)$ can be calculated to extreme accuracy for short times, with virtually no scatter, to $t$ well beyond $\tau_{RW}$. Applying this calibration to the SISF, the transit motion is a random walk of step rate $\nu$ and step distance $\delta{R}$, beginning at $\tau_{RW}$. The damping factor is given by equation~(\ref{eq17}), so that
\begin{equation} \label{eq20}
F_{VT}^{s}(q,t)=F_{VT}^{s}(q,\tau_{RW}) \;\;e^{-\gamma(q\;\delta R)(t-\tau_{RW})}       , \;\;   t\geq \tau_{RW}.
\end{equation}
The factor $F_{VT}^{s}(q,\tau_{RW})$ is given by equation~(\ref{eq19}) evaluated at $t=\tau_{RW}$. Comparison of equation~(\ref{eq20}) with $F_{MD}^{s}(q,t)$ in figure~\ref{fig6} shows excellent agreement at all $t$ beyond $\tau_{RW}$. Just as for the MSD, the theory is rather insensitive to the choice of  $\tau_{RW}$. The reason is evident from figure~\ref{fig5}, which shows the crossover theory and MD nearly parallel for a while beyond $\tau_{RW}$. The steady-state random walk produces exponential damping for $t\geq\tau_{RW}$.

\section{Extension to all $q$}

\begin{figure} [h]
\includegraphics [width=0.70\textwidth]{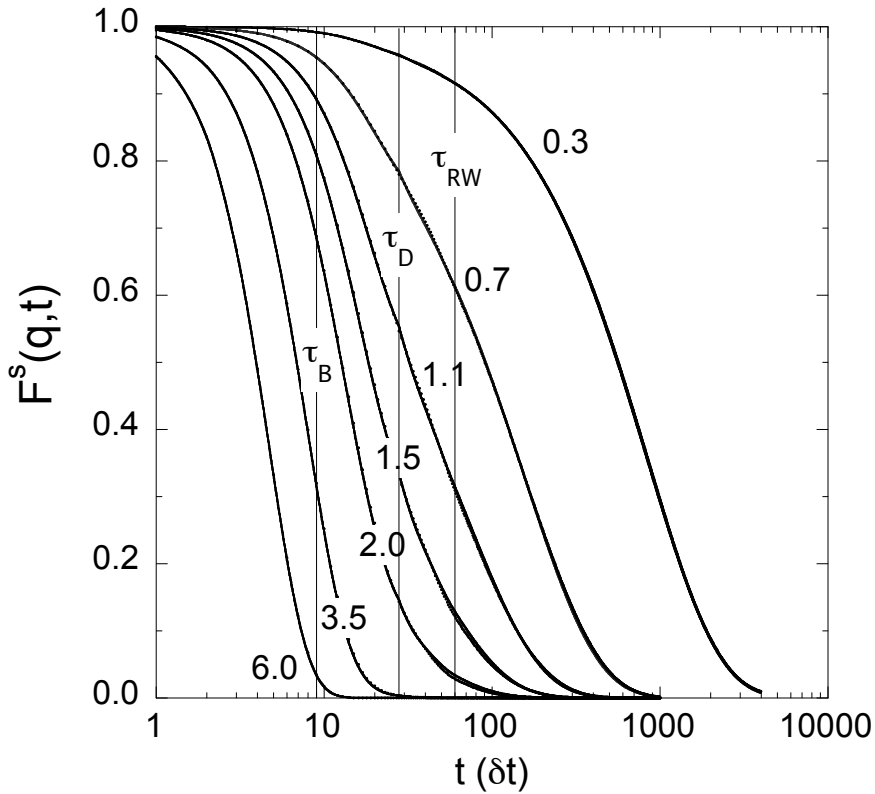}
\caption{For $q$ from $0.3$ to $6.0$  a$_{0}^{-1}$: Dots are $F_{MD}^{s}(q,t)$ and lines are $F_{VT}^{s}(q,t)$, where the $t$ scale is logarithmic. Deviation is so small it can be detected only at short segments on a few curves.}
\label{fig7}
\end{figure}

We shall now apply the theory to a wide range of $q$, a range for which the initial structural correlation $F_{vib}^{s}(q,\infty)$ varies from near one at small $q$ to zero at large $q$. Comparison of $F_{VT}^{s}(q,t)$ with $F_{MD}^{s}(q,t)$ for seven $q$ is shown in figure~\ref{fig7}. The deviation function (the error) is 
$\Delta{F^{s}(q,t)}$, defined by
\begin{equation} \label{eq22}
\Delta F^{s}(q,t)=F_{VT}^{s}(q,t) - F_{MD}^{s}(q,t). 
\end{equation}
The deviation and other relevant parameters are listed for eight $q$ in Table \ref{Table2}. The accurate value of $q_1$ is  $1.1050$  a$_{0}^{-1}$.
\begin{table}[ht] 
\caption{\label{table2} Analysis of the deviation function. $F_{vib}^{s}(q,\infty)$ measures the structural correlation at $t=0$. $\gamma(q\delta R)$ is the transit damping coefficient on the random walk interval, equation~(\ref{eq20}). $t_{\epsilon}(q)$ is when $F_{MD}^{s}(q,t)$ reaches $\epsilon=0.002$, our numerical measure of zero. Columns 5-7 respectively list the largest deviation of the negative dip (vibrational interval), spread of deviation (crossover interval), and largest deviation of the negative dip (random walk interval).}
\begin{ruledtabular}
\begin{tabular}{ccccccc}
$q$(a$_{0}^{-1})$&$F_{vib}^{s}(q,\infty)$&$\gamma (q\delta R) $(ps$^{-1})$&$t_{\epsilon}(q)$(ps)&$0\leq t \leq \tau_{D}$&$\tau_{D}\leq t \leq \tau_{RW}$& $t \geq\tau_{RW}$\\
\hline
0.29711&0.94185&0.1733&30   & 0.000&-0.001 to 0.000&-0.003\\
0.70726&0.71699&0.9222&7.0  &-0.002&-0.005 to 0.004&-0.005\\
  1.1050&0.45332&2.0146&3.1  &-0.003&-0.006 to 0.006&-0.003\\
  1.5052&0.24047&3.1805&1.8  &-0.006&-0.004 to 0.007&-0.001\\
  2.0041&0.08776&4.2975&1.0  &-0.007&-0.002 to 0.005&0.000\\
  2.5064&0.02573&4.7423&0.60&-0.006& 0.000 to 0.003&0.000\\
  3.5008&0.00126&3.9994&0.22&-0.003& 0.000 to 0.000&0.000\\
  6.0013&0.00000&4.2271&0.08& 0.000& 0.000 to 0.000&0.000\\
\end {tabular}
\end{ruledtabular}
\label{Table2}
\end{table}

We shall describe the observable deviation features and assess their significance. The deviation at each $q$ exhibits a negative dip during the vibrational interval, which is visible in figure~\ref{fig4} and whose maximum negative deviation is listed in Table II. The likely cause of this dip is a small negative transit  contribution omitted from equation~(\ref{eq18}); however, something like finite-$N$ error cannot be ruled out.

In fitting a random walk to the MSD crossover, and applying the same approximation to the SISF crossover, we are neglecting explicit nonlinear $t$ dependence of the transit motion. The corresponding errors in Table II are negligible, but this need not always be the case. We expect the crossover to show more complex behavior with increasing $T$.

Finally, the deviation exhibits a negative dip during the random walk interval for small $q$. The largest magnitude is at $q=0.70726$a$_{0}^{-1}$, Table II, and lies at $t$ around $400 \delta{t}$ in figure~\ref{fig7}. Our overall assessment of Table II is as follows. At all $q$ and $t$, the deviation is small enough   that it indicates no significant error, numerical or theoretical, and no attempt to refine the numerical accuracy is warranted at this time.

In our original study of the SISF~\cite{G1}, the crossover interval was not resolved and theory was based on two time intervals, vibrational and random walk. The self dynamic structure factor was shown moderately more accurate then benchmark mode coupling theories~\cite{LV1970,WS1982}. In the present formulation, $\Delta(q,t)$ is everywhere smaller than the original by a factor of $3$ or more.

The original formulation is still correct, accurate, and a source of useful analysis. Two important limits were shown by analysis to derive   from V-T theory~\cite{G1}, the hydrodynamic limit at small $q$ and the free particle limit at  large $q$. Table II shows the vanishing of deviations as $q$ decreases toward the hydrodynamic regime. On the other hand, the free particle limit provides information not previously available for  liquids on the overall convergence to the large $q$ limit~\cite{G1}. Notice the free particle motion is the leading $t$ dependence of vibrational theory, equation~(\ref{eq7}). Notice also as $q$ increases, 
 $F_{vib}^{s}(q,\infty)$ decreases to zero, figure~\ref{fig2}. When we follow $F_{VT}^{s}(q,t)$ as $q$ increases from the hydrodynamic regime, say via figure~\ref{fig7}, we see that $F_{VT}^{s}(q,t)$ goes to zero at an ever decreasing time; this time decreases through $\tau_{RW}$, then through $\tau_{D}$, and ultimately through $\tau_{B}$. The damping process changes dramatically as $q$ increases, but the initial motion is always vibrational and its initial segment is always free particle. This behavior is shown for both  $F_{VT}^{s}(q,t)$  and its Fourier transform  $S_{VT}^{s}(q,\omega)$ in figures~ (6)-(11) of~\cite{G1}. It is also shown in Table~II as the vanishing of the deviation as $q$ increases toward the free particle regime.
 
 \section{Comparison of theories}

We begin by comparing the goal and the operational techniques of V-T theory with those of the broader field of research in dynamics of noncrystalline materials. In the past few decades, the broader field has focused on understanding the glass transition and the glassy state, and has coalesced around the technique of mode coupling theory (MCT). This technique has the remarkable ability to treat simple and complex systems with equal ease, from monatomic to molecular liquids, molten salts, colloids, polymers, and large organic molecules in the liquid phase. Diverse and valuable reviews are available~\cite{Gotze1999, Das2004, BK2005, BB2011}.

V-T theory specifically addresses liquid dynamics, and applies to equilibrium liquids at all temperatures, allowing comparison with MCT studies of equilibrium supercooled liquids. On the face of it, however, that comparison is not straightforward. For a given time correlation function, MCT evaluates the contributions to a generalized Langevin equation, while V-T theory evaluates the function's defining equation. The working measures of the atomic motion are quite different in the two theories. But the actual atomic motion is unique, and we can base a comparison of theories on their respective descriptions of that motion.

\begin{figure} [h]
\includegraphics [width=0.70\textwidth]{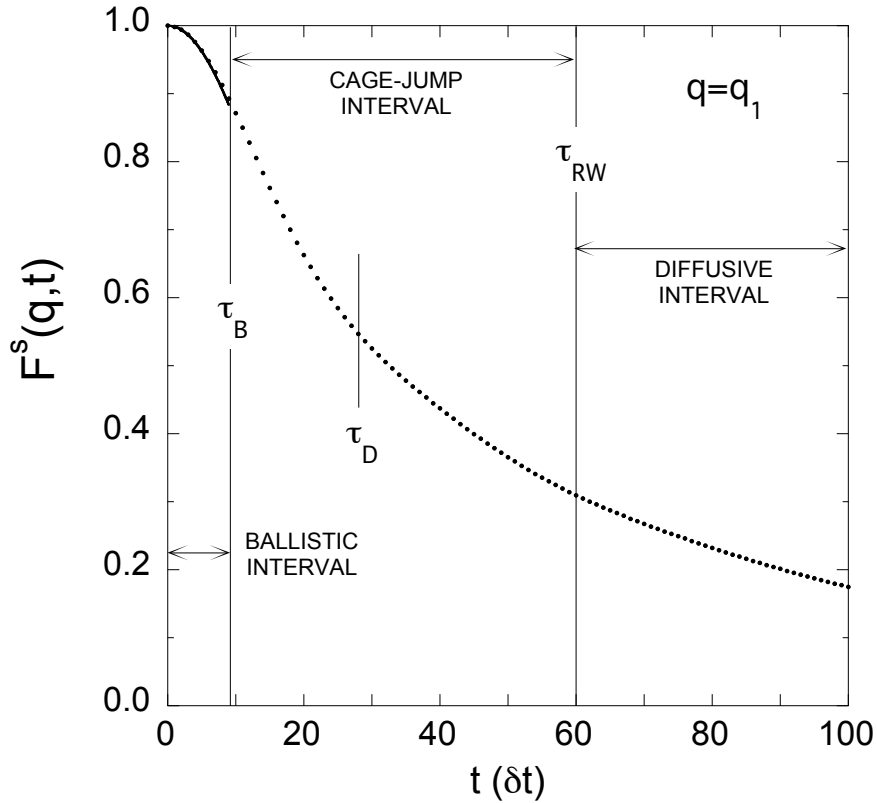}
\caption{The time evolution commonly observed in MD data for the MSD in normal and supercooled liquids, shown for the SISF of liquid Na. Dots are $F_{MD}^{s}(q,t)$ and short line to $\tau_{B}$ is the ballistic contribution.}
\label{fig8}
\end{figure}

For normal and supercooled liquids, MD data for the MSD time evolution~\cite{KA1995} has a well established interpretation in terms of the atomic motion~\cite{Kob1999}. Here we apply the same interpretation to the SISF, as shown in figure~\ref{fig8}. There are three successive time intervals, characterized respectively by ballistic, ``cage-jump'' and diffusive motions. Upon supercooling the liquid, the primary change in figure~\ref{fig8} is the appearence of a plateau in the middle interval, where the atomic dynamics is slowed by cage motion. G{\"o}tze tells us MCT for density fluctuation dynamics was developed originally to deal with the cage effect (section 1 of ~\cite{Gotze1999}. The  idea is that a particle remains trapped for a while in the cage of its neighbors. The process of exiting a cage is a jump, and cage-jump motion characterizes the middle interval. The time evolution  of figure~\ref{fig8} is commonly observed, as for example in binary LJ and silica systems (figure 3 of~\cite{Kob1999}, complex hard sphere systems~\cite{Toku2008}, a one-component LJ system~\cite{Hoang2013}, and Al~\cite{JP2013}.

Figure~\ref{fig8} compares directly with the V-T time evolution in figure~\ref{fig3}. In the comparison  we can see two theoretical advantages in working with vibrational motion in place of the ballistic motion: vibrational motion maintains agreement with MD through a much greater amount of damping, until $\tau_{D}$, and the vibrational contribution goes on to saturate near the level of zero vibrational correlation. On the other hand, these theoretical advantages can be lost in complicated systems, due to the increased complexity of the vibrational Hamiltonian.  Finally in comparing figures~\ref{fig8} and \ref{fig3}, we note that the two plans are the same on $0 \leq t \leq \tau_{B}$; are the same on $t \geq \tau_{RW}$ when the diffusive motion is represented by the steady state transit random walk; and that cage-jump and vibration-transit motions have a logical correspondence on $\tau_{B} \leq t \leq \tau_{RW}$.

The motional correspondence just mentioned is apparent in the literature. Cage breaking processes in binary LJ systems can be related to long term diffusion~\cite{dSW2008}, and can be modeled by a correlated random walk~\cite{dSW2009}. This description matches transits in monatomic liquids, where random walk correlations are minimal because all atoms are equivalent~\cite{G1}. It has been found that atoms in a binary LJ system involve the sensible displacement of a small group of particles, and that these jumps are diffusive at high $T$, subdiffusive at low $T$~\cite{CPC2016}. The transits observed in Na and Ar also occur in correlated groups, but are diffusive at all $T$~\cite{WCC2001}. Models which connect fast and slow degrees of freedom of viscous liquids have been discussed~\cite{Dyre2006}, and distinct fast and slow channels in a LJ fluid have been found~\cite{BNGBB2015}. The fast-slow connection in V-T theory is the three-interval evolution upon which the present work is based, from the initial pure vibrational motion (fast) to the ultimate pure transit motion (slow) (see also ~\cite{MSD}). Finally, an example of the formidable complications MCT is able to address is a study of flexible trimers rattling in the cages of their neighbors~\cite{BL2016}.

\section{Conclusions}

In this paper and the previous one~\cite{MSD}, V-T theory has demonstrated its analytical tractability and mathematical simplicity. In this paper, the theory with \emph{apriori} vibrational calibration, and with transit calibration from the MSD, accounts for the SISF to extreme accuracy at all $q$ and $t$. Two questions lead us to deeper insight into what all this means.

(a) Considering the massive $q$ dependence of $F_{MD}^{s}(q,t)$, figure~\ref{fig7}, how can we have a $q$-independent calibration? It is because we calibrate the motion, not the time correlation function. The atomic motion has no $q$ dependence. Vibrational motion is given by the vibrational Hamiltonian, and transit motion is given by  parameters calibrated from MD (Table I). All $q$ dependence is in equation~(\ref{eq1}), and carries over to the vibrational time correlation functions and Debye-Waller factors, equations~(\ref{eq4})-(\ref{eq6}), and to the transit damping coefficients of  equations~(\ref{eq19}) and (\ref{eq20}).

(b) What makes the theory uniformly accurate over all $q$ and $t$? With reference to figure~\ref{fig7}, the answer has two parts. (i)  The vibrational motion is accurate because it is calculated from one of the liquid's own $3N$-dimensional random valleys, and the liquid measures only vibrational motion at $t\leq \tau_{D}$ and all $q$.  (ii) The transit motion is accurate because the steady-state transit random walk is calibrated from the MSD measurement of $D$, and the liquid measures only random walk motion at $t\geq \tau_{RW}$ and all $q$. This leaves the crossover interval accurately calibrated at its end points at all $q$.

A common three interval time evolution plan, figure~\ref{fig8}, can be applied to the MSD and SISF, in MCT or V-T theory. The atomic motion for the two theories is either the same or logically equivalent on each interval (section VI). 

We are currently applying the present theory to critical slowing down in supercooled liquid Na.

\acknowledgments{We are pleased to thank Brad Clements for helpful and encouraging discussions. This research is supported by the Department of Energy under Contract No. DE-AC52-06NA25396.}

\bibliography{LiquidTheory}

\end{document}